\documentclass[aps,prb,twocolumn,floatfix,superscriptaddress]{revtex4-2}
\usepackage{graphicx}
\usepackage{amsmath,amsthm}
\usepackage{amssymb}
\usepackage{braket}
\usepackage{xcolor}
\usepackage{bm}
\usepackage{ulem}
\usepackage[bookmarks=false,linkcolor=blue,urlcolor=blue,colorlinks,citecolor=blue]{hyperref}
\usepackage{caption}
\usepackage{subcaption}
\usepackage[skip=10pt plus1pt, indent=40pt]{parskip}

\newcommand{\bea}{\begin{eqnarray}}
	\newcommand{\eea}{\end{eqnarray}}
\newcommand{\beq}{\begin{equation}}
	\newcommand{\eeq}{\end{equation}}

\newcommand{\varsha}[1]{{\color{black} #1}}

\begin{document}

\title{Geometry contribution to sound attenuation in double-Weyl semimetals}

\author{Varsha Subramanyan}
\email{varshas@sjtu.edu.cn}
\altaffiliation{Present address: Tsung-Dao Lee Institute, Shanghai Jiao Tong University, Shanghai, China}
\affiliation{Theoretical Division, T-4, Los Alamos National Laboratory, Los Alamos, New Mexico 87545, USA}

\author{Shi-Zeng Lin}
\email{szl@lanl.gov}
\affiliation{Theoretical Division, T-4, Los Alamos National Laboratory, Los Alamos, New Mexico 87545, USA}
\affiliation{Center for Nonlinear Studies (CNLS), Los Alamos National Laboratory, Los Alamos, New Mexico 87545, USA}
\affiliation{Center for Integrated Nanotechnologies (CINT), Los Alamos National Laboratory, Los Alamos, New Mexico 87545, USA}
\author{Avadh Saxena}
\email{avadh@lanl.gov}
\affiliation{Theoretical Division, T-4, Los Alamos National Laboratory, Los Alamos, New Mexico 87545, USA}
\affiliation{Center for Nonlinear Studies (CNLS), Los Alamos National Laboratory, Los Alamos, New Mexico 87545, USA}

\begin{abstract}
The axial coupling of strain to the nodes of the simple Weyl semimetals leads to anomalous contributions to sound attenuation in such materials. However, in double Weyl semimetals, there is no such axial coupling. Strain instead couples as a symmetry-breaking director field that deforms the Fermi surface around each Weyl node. In this work, we show that absence of axial coupling in double Weyl semimetals implies a very different mechanism of relaxation due to sound. The deformed geometry of the Fermi surface is the only source of sound attenuation under these conditions. Thus, we identify a geometric contribution to sound attenuation in double Weyl semimetals that is entirely absent in simple Weyl semimetals. 
\end{abstract}
                                                        
\maketitle
\section{Introduction}

The spectrum of Weyl semimetals (WSMs) is distinct from that of normal metals/semimetals, valley semiconductors, etc. because they are endowed with topology and a chiral channel that connects the Weyl points to each other. \cite{PhysRevB.83.205101,Lu_Wang_Ye_Ran_Fu_Joannopoulos_Soljačić_2015,Xu_Belopolski_Sanchez_Zhang2015,RMP, TaAs, AR,Yan_Felser_2017}. The chiral channel leads to several anomalous transport signatures that serve as material signatures of topology as well as realizations of the chiral anomaly\cite{SciPostPhysLectNotes.62}. An important consequence of the topological nature of Weyl fermions is that lattice deformations couple to the low-energy electrons in unconventional ways. In simple Weyl semimetals, strain shifts the two Weyl nodes in opposite directions in momentum space and therefore acts as an axial (pseudo-)gauge field that couples minimally to the low-energy Hamiltonian. \cite{Phonons-and-electron-phonon,Symmetry-based,VOZMEDIANO2010109,Elastic_Gauge,PhysRevLett.134.026304,Pikulin,Inhomogeneous-Weyl-and-Dirac-Semimetals,Phonon_Helicity,Hall-viscosity-for-optical-phonons,PhysRevB.111.035126} More generally, however, strain can generate a broader class of emergent fields beyond gauge fields. Depending on the crystal symmetry and the nature of the deformation, it can also induce emergent frame (vielbein) fields that modify the effective geometry experienced by the quasiparticles and give rise to geometric responses. \cite{PhysRevLett.108.227205,PhysRevB.102.235163,Chen_Zhang_Cao_Lin_Xiao_2025} These emergent gauge and geometric fields provide a unified description of how lattice deformations influence topological fermions. The emergent gauge and geometric fields generated by strain lead to a variety of unconventional transport phenomena, including anomaly-induced magnetotransport, pseudomagnetic-field effects, ultrasonic attenuation, and nonlocal transport, which have been extensively investigated theoretically and experimentally. \cite{Sukhachov,Antebi,Ilan2020PseudoFields,Hu2019Transport}

This intuition, however, no longer applies to multi-Weyl semimetals, whose Weyl nodes carry a topological charge (winding number) $|n|>1$. Unlike simple Weyl nodes with linear dispersion in all momentum directions, multi-Weyl nodes exhibit anisotropic low-energy dispersions protected by crystalline rotational symmetry, leading to quadratic ($n=2$) or cubic ($n=3$) band crossings in the plane transverse to the symmetry axis. Candidate materials include the magnetic spinel HgCr$_2$Se$_4$ and SrSi$_2$. \cite{material1,material2,Singh2018} In previous work \cite{OldPaper}, we highlighted that strain has a very different nature when coupled to multi-Weyl semimetals as compared to simple Weyl semimetals. In these materials, strain breaks the isotropy of the Fermi surface and drives a nematic transition. The higher winding number Weyl nodes also split into simple Weyl nodes with winding $\pm 1$ since the rotational symmetry protecting them has been broken by strain. The extent of deformation of the Fermi surface is captured to leading order by the quantum geometric tensor\cite{PhysRevLett.131.240001} and is reflected in strain-induced modifications to the conductance of the multi-WSMs. 

In this work, we investigate the effects of dynamic strain (i.e., sound waves) on multi-WSMs with $n=2$, with emphasis on contributions of the quantum geometric tensor. We study the mechanism of attenuation of sound waves in these materials in contrast to that in simple WSMs. In simple WSMs, the axial coupling of strain results in an effective pseudomagnetic field that acts on each Weyl node with an opposite sign. This differential coupling contributes an anomalous mode of relaxation to the attenuation of sound waves. However, the coupling of strain to multi-Weyl semimetals is neither axial nor gauge-like. This results in the anomalous contribution vanishing in double-Weyl semimetals as well as over materials of {\it even} winding.  Instead, the modification of the geometry of the Fermi surface due to strain results in a geometric contribution to the attenuation of sound waves. In materials with {\it odd} winding, both mechanisms contribute to sound attenuation. 

To estimate the amount of sound attenuation in multi-Weyl semimetals, we evaluate the coefficient of energy attenuation \cite{AAA} when a “slow" sound wave $\vec{u}=\vec{u}_0e^{i\vec{q}.\vec{x}-i\omega t}$ passes through the material. In terms of the rate of change of band energy $Q$, intensity of the sound wave $I$ and volume of the material $V$, the coefficient of energy attenuation is given by 
\begin{align}
    \Gamma=\frac{Q}{VI},\quad\quad
    I=\rho \langle|\partial_t\vec{u}|^2\rangle=\frac{1}{2}\rho \omega^2u_0^2 \,, 
\end{align}
where $\rho$ is mass density of the material. The quantity Q (rate of change of energy) is given by
\begin{align}
    Q=\langle\partial_t E\rangle
    =-i\omega V\int\frac{d^3k}{(2\pi)^3} E f(E-\mu) \,, 
\end{align}
and can be evaluated through chiral kinetic theory and Boltzmann transport equations \cite{CKT}. We outline the mechanism of sound attenuation in simple Weyl semimetals (as described in earlier works) in Section \ref{sw}, and then show in Section \ref{dsw} how the mechanism is entirely different in double-Weyl semimetals. We analyze the regimes of applicability of our results in Section \ref{ra} and end with a brief discussion and outlook in Section \ref{disc}.

\section{Simple Weyl Semimetals}\label{sw}

\begin{figure}
    \centering
    \includegraphics[width=0.45\textwidth]{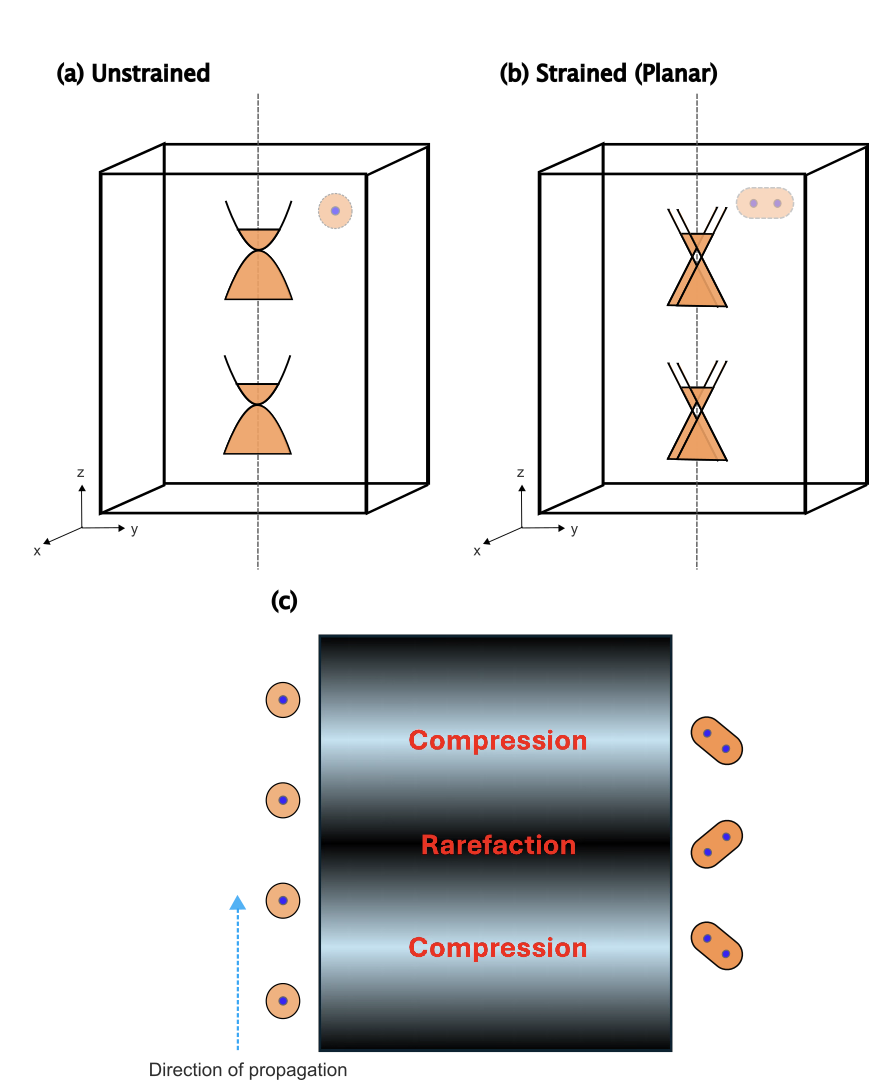}
    \caption{Pictorial representation of the change in Fermi surface geometry due to a sound wave: \varsha{This figure shows the low-energy dispersion of a double-Weyl semimetal and the shape of the Fermi surface (insets) for a chemical potential close to the nodes for a (a) unstrained and (b) strained system. The blue dots in the Fermi surface insets indicate the location of the Weyl node. (c) When a slow sound wave passes through this system, the material varies periodically between these two states. When applied strain due to the sound wave is zero between a compression and rarefaction, the Fermi surface is isotropic again as seen on the left of sound wave. However at other times in the cycle, the Fermi surface is nematic in shape as seen on the right side. The change in sign from a compression to a rarefaction is reflected in a 90 degree relative angle between the direction of splitting of the Weyl-nodes.}}
    \label{fig1}
\end{figure}

The essentials of sound attenuation in simple Weyl semimetals are summarized here for completeness \cite{Sukhachov,Antebi}. As mentioned before, the strain couples like a pseudo-gauge field to this system. However, these vector-like perturbations do not cause significant sound attenuation. Instead, the scalar component of strain coupling is the term of interest. Following the convention of Bardeen, we can write the modified band structure as
\begin{align}
    E_s=E_0+\lambda_{ij}u_{ij}=E_0+(\lambda_{ij}^e+s\lambda_{ij}^o)u_{ij} \,, 
\end{align}
where $\lambda_{ij}$ are coupling constants known as the deformation potential. It generally has odd and even components $\lambda_{ij}^o$ and $\lambda_{ij}^e$, respectively. The multiplier $s$ indicates the sign of the Weyl node. The rate of change of energy is thus given by 
\begin{align}
\begin{split}
    Q&=-i\omega V\sum_s\int\frac{d^3k}{(2\pi)^3} \lambda_{ij}u_{ij} f_s(E_s-\mu_s)\\&=-i\omega V \sum_s \lambda_{ij}u_{ij}\bar{f}_s(\mu_s) \,. 
    \end{split}
\end{align}

That is, Q depends entirely on the Fermi surface average of the modified distribution function, which is generally affected by strain, external EM fields, the chiral anomaly as well as the frequency of the sound wave.

%This quantity can be evaluated by considering the Boltzmann equation again:
%\begin{align*}

    %\partial_t f_++& \nabla_{\bf x}f_+.{\bf \dot{x}} +\nabla_{\bf k}f_+.{\bf \dot{k}} = -\frac{f_+-\bar{f_+}}{\tau_{intra}}-\frac{f_+-\bar{f_-}}{\tau_{inter}}\\&=-(f_+-\bar{f_+})\bigg(\frac{1}{\tau_{inter}}+\frac{1}{\tau_{intra}}\bigg)-\frac{\bar{f_+}-\bar{f_-}}{\tau_{inter}}
%\end{align*}
%where the overhead bar indicates Fermi surface average. If we average both sides of the equation over the Fermi surface, we obtain
%\begin{align}
%    -\frac{\bar{f_+}-\bar{f_-}}{\tau_{inter}}=\overline{\partial_t f_++ \nabla_{\bf x}f_+.{\bf \dot{x}} +\nabla_{\bf k}f_+.{\bf \dot{k}}}
%\end{align}
%We will call this term on the LHS $\Lambda$ for convenience. A second equation is obtained using the Gauss law:
%\begin{align}
%    {\bf \nabla.E}=\bar{f_+}+\bar{f_-}
%\end{align}
%We can solve these equations simultaneously after plugging in the equations of motion for $\dot{x}_i$ and $\dot{k}_i$. Depending on the kinds of external fields present as well as the relative magnitude of the various time scales involved ($\tau_{inter}, \tau_{intra}$ and $\omega^{-1}$), we can obtain several possible expressions for Q, and hence $\Gamma$. It is useful to look at one representative form for $\Gamma$.

We direct the reader to Ref. \cite{Sukhachov} where this quantity is elegantly estimated for simple Weyl semimetals. Here we simply quote the expression in a relevant frequency regime.  In the presence of an external magnetic field and in the regime $\omega \tau_{eff}\ll 1$, we obtain
\begin{align}
    \Gamma&=\frac{2\nu(\mu_s)|\lambda_{ij}^o|^2}{\rho_m v_s}\frac{q^2}{1/\tau_{eff}+({\bf v_s.\hat{q}})^2/D} \,, \\
    1/\tau_{eff}&=1/\tau_{inter}+q^2D \,, 
\end{align}
where ${\bf v_s}\sim s{\bf B}/\nu(\mu_s)$, $D=v_F^2\tau_{intra}/3$, \varsha{$\tau_{inter}$ is the inter-node relaxation rate, $\tau_{intra}$ is the intra-node relaxation rate }and $\nu(\mu)$ is the density of states. The key thing to notice here is that $\Gamma \sim|\lambda_{ij}^o|^2\omega^2\tau_{eff}$, the odd component of the scalar coupling of strain/sound. This is true for every estimation of $\Gamma$ in every regime of applicability. The fact that the odd/axial term is present but the even term $\lambda_{ij}^e$ is not present is a consequence of the axial coupling of strain to simple Weyl semimetals due to inversion symmetry. It is this factor that is different in double-Weyl semimetals, we shall discern in the next section.

\section{Double-Weyl Semimetals}\label{dsw}

The crucial difference in the analysis of double-Weyl semimetals in the presence of sound comes from the way in which the energy bands are modified by the presence of strain \cite{OldPaper}. A detailed presentation of the effect of strain on double-WSMs is presented in Ref. \onlinecite{OldPaper}. In summary, the minimal low-energy Hamiltonian describing a double-WSM is given by
\begin{align}
    H&=v_1a^2(k_x^2-k_y^2)\sigma_x+v_1a^22k_xk_y\sigma_y+v_2(k_za\mp\frac{\pi}{2})\sigma_z \,, 
\end{align}
\varsha{where $a$ is the lattice constant}. In the presence of lattice displacement \varsha{${\bf u}({\bf x})$}, strain is described by the strain tensor $u_{ij}=\frac{1}{2}(\partial_iu_j+\partial_ju_i)$. The low-energy Hamiltonian is modified to 
\begin{align}
    \begin{split}
    H&=[(k_x^2-k_y^2)-\gamma(u_{xx}-u_{yy})]\sigma_x\\&+(2k_xk_y-2\gamma u_{xy})\sigma_y+[(k_z\mp\frac{\pi}{2} \mp\gamma\frac{\pi}{2} u_{zz})]\sigma_z \,, \label{sth} 
\end{split}
\end{align}
where $\gamma$ is the Gr\"uneisen parameter. \varsha{We have ignored dimensionful constants/energy scales for convenience. We will restore them later.} As in the case of simple WSMs, strain is a vector perturbation given by ${\bf C}=\gamma\bigg((u_{xx}-u_{yy}),\ 2 u_{xy},\ \frac{\pi}{2} u_{zz} - (u_{xx}+u_{yy})\bigg)$. However, strain is not a pseudo-gauge field in these materials as seen in the absence of minimal coupling. Instead, strain breaks the $C_4$ symmetry of the original Hamiltonian, thus splitting the $\pm 2$ Weyl nodes to $\pm 1$ \cite{OldPaper}. The Fermi surface also loses its isotropy and becomes nematic. A pictorial representation of this process is shown in Figure \ref{fig1}(a) and 1(b). Further details on how to derive the low-energy Hamiltonian are furnished in Appendix \ref{A}. In the rest of this work, we focus on the effects of strain in the planar section and ignore the $k_z$ direction without loss of generality. Including the $k_z$ direction offers no qualitative changes to the following analysis. 
%, and a brief discussion of the same is given in Appendix \ref{B}.

Again following the Bardeen convention, the strain-induced modification of the energy bands is given by
\begin{align}
\begin{split}
    E&=E_0+\delta E = E_0 -\frac{(k_x^2-k_y^2)}{E}C_x-\frac{2k_xk_y}{E}C_y\\&=E_0 -\lambda_{ij}({\bf k})u_{ij} \,. 
\end{split}
\end{align}
That is, the elements of the deformation potential are explicitly dependent on momentum. By inversion symmetry, it is also clear that the perturbation term $\lambda_{ij}u_{ij}$ has no odd terms that flip sign from one Weyl node to the other, since the momentum terms are all quadratic. Therefore, there are no terms analogous to $\lambda_{ij}^o$. This would be true even in the presence of scalar coupling terms by the same argument. Thus, the mechanism of sound attenuation seen in simple Weyl semimetals would be entirely absent in double-Weyl semimetals. We therefore need to evaluate this system in a different way.

Further, in multi-WSM, $\delta E$ is dependent on the quantum geometric tensor $g_{ij}$ associated with the Weyl node. It is explicitly given by 
\begin{align}
\begin{split}
    \delta &E=-\bigg(\frac{C_x}{E_0}\frac{(g_{xx}-g_{yy})}{W_2^2}+\frac{C_y}{E_0}\frac{2g_{xy}}{W_2^2}\bigg)\\
    &=-\bigg(\frac{(u_{xx}-u_{yy})}{E_0}\frac{(g_{xx}-g_{yy})}{W_2^2}+\frac{2u_{xy}}{E_0}\frac{2g_{xy}}{W_2^2}\bigg),
    \end{split}
\end{align}
where $W_2=2v_1^2\frac{(k_x^2+k_y^2)}{E_0^2}$ is a term proportional to the density of states at each node. Expressing the band structure modification in this manner emphasizes how the change in energy is a coupling between covariant terms of the spatial geometry (the strain tensor) and momentum space geometry (quantum geometric tensor). In the presence of dynamic strain, or a sound wave, the Weyl nodes oscillate between split and unsplit, leading the Fermi surface to oscillate between isotropic and nematic as shown in Figure \ref{fig1}(c). This oscillating Fermi surface is absent in simple WSMs and thus, offers no contribution to sound attenuation. However, we will demonstrate that this change in geometry of the Fermi surface is the most important contribution to sound attenuation in double-WSMs. 

\subsection{Rate of change of energy and the Boltzmann equation}

In order to calculate the rate of change of energy $Q=\langle\dot{E}\rangle=-i\omega\langle\delta E\rangle$ in this case, we need to evaluate the modified electron distribution function. It is given by
\begin{align}
\begin{split}   
Q&=-i\omega V\int\frac{d^3k}{(2\pi)^3}\delta E f(E-\mu)\\&=i\omega V\int\frac{d^3k}{(2\pi)^3} f(E-\mu) \bigg(\frac{C_x}{E}(k_x^2-k_y^2)+\frac{C_y}{E}2k_xk_y\bigg) \,. 
\end{split}
\end{align}
On writing the modified electron distribution at a particular Weyl node as $f(E)=\bar{f}(E,C_i) +\delta f(k_i,C_i)$, that is, as a sum of the average distribution function plus deviations from the average, it is clear that contributions from the Fermi surface average are identically zero, irrespective of the nature of $\bar{f}$ since the terms $k_x^2-k_y^2$ and $2k_xk_y$ will always yield zero over the Brillouin zone. Therefore, only deviations from average occupancy of the energy levels contribute to the geometric part of sound attenuation. Thus, we have
\begin{align}
\begin{split}   
    Q&=i\omega V\int\frac{d^3k}{(2\pi)^3} \delta f(E-\mu) \bigg(\frac{C_x}{E}(k_x^2-k_y^2)+\frac{C_y}{E}2k_xk_y\bigg) \,, \\ \label{int}
    \delta f&= \tau^*\partial_Ef_0 (\dot{E}+\dot{x}_i\partial_{x_i}E+\dot{k}_i\partial_{k_i}E) \,, 
    \end{split}
\end{align}
where we have used the Boltzmann transport equation to estimate $\delta f$, using $\tau^*$ as the effective relaxation rate analogous to $\tau_{eff}$ defined in the previous section. The term $f_0$ is the unperturbed electron distribution function, assumed to be $\theta(\mu-E)$ at low temperatures. On plugging the Boltzmann equation back into the expression for $Q$, it is seen that there is an obvious leading contribution due to the $\dot{E}$ term. That is,
\begin{align}
\begin{split}   
    Q&=-\tau^*\omega^2 V \sum_{s=\pm}\int\frac{d^3k}{(2\pi)^3}(\delta E)^2 \partial_Ef_0(E-\mu_s)\\&=-\sum_{s=\pm}V\frac{\tau^*\omega^2\mu_s}{32\pi}(C_x^2+C_y^2)\label{Q} \,, 
\end{split}
\end{align}
\varsha{where $\mu_s$ is the chemical potential at each node.}
To evaluate the other terms, we use the modified chiral kinetic theory developed in Refs. \onlinecite{OldPaper} and \onlinecite{CKT}. A more detailed explanation of terms and notations is given in Appendix \ref{EOM}. The equations of motion in the presence of strain are given by
\begin{align}
    \dot{x}_i(\delta_{ij}+S^T_{ij})&=v_j+({\bf \dot{k}}\times {\bf\Omega})_j \,, \\
    \dot{k}_i(\delta_{ij}+S_{ij})&=E^t_j+({\bf \dot{x}}\times {\bf B^t})_j \,, 
\end{align}
where ${\bf E^t}={\bf \tilde E}-{\bf \nabla_x}E+\partial_t{\bf A^x}$ and ${\bf B^t}={\bf B}-{\bf \nabla\times A^k}$  are the total modified fields. {\varsha{Here, ${\bf \tilde E}$ is the external electric field and ${\bf B}$ is the external magnetic field.}} We also have cross-terms of the generalized Berry curvature 
\begin{align}
    S_{ij}=\partial_{k_j}A^x_i-\partial_{x_i}A^k_j \,, 
\end{align} where $A^x_i$ and $A^k_i$ are spatial (strain-induced) and momentum Berry connections over phase space, respectively.  

\varsha{As is the norm for estimating transport through the Boltzmann method, we now solve for the equations of motion $\dot{x}_i$ and $\dot{k_i}$. The presence of the $S_{ij}$ terms makes the equations of motion deviate from those obtained for any quantum anomalous Hall system. Details of the equations of motion and how to evaluate them are given in Appendix \ref{EOM}. They are then plugged into Eq. (\ref{int}) to evaluate $Q$ term by term. However,} after some routine (though involved) algebra, each term in the integral for $Q$ vanishes identically in the limit $\omega\tau^*\ll 1$. In this limit, the presence of the $k_x^2-k_y^2$ and $2k_xk_y$ terms in the integral in Eq. (\ref{int}) makes most terms vanish on integrating them over the Brillouin zone. Other terms vanish because they are proportional to $(\partial_xC_x\partial_yC_y-\partial_xC_y\partial_yC_x)$, which vanishes  when we consider that this term equals $C_xC_y(q_xq_y-q_xq_y)$. Therefore, the first term we evaluated in Eq. (\ref{Q}) is the only contribution. This term is quadratic in strain, does not depend on EM fields and is a geometric contribution.  When divided by intensity $I=\frac{1}{2}\rho \langle|\partial_t\vec{u}|^2\rangle=\frac{1}{2}\rho \omega^2u_0^2$ to obtain $\Gamma$, we get 
\begin{align}
    \Gamma=V\frac{v_1}{v_2}\frac{\tau^*\omega^2\mu}{16\pi a^3}(C_x^2+C_y^2)/\frac{1}{2}\rho \omega^2u_0^2V=\frac{\tau^*\omega^2\mu}{8\pi a^3}\frac{v_1}{v_2}\frac{\gamma^2}{\rho v_s^2},
\end{align}
\varsha{where we have set the chemical potential of each Weyl node to be equal $\mu_+=\mu_-=\mu$ and restored dimensionful constants in the final expression. This is the main result of our work. It is notable that rate of change of energy $Q$ has no magnetic field dependence up to second order in strain. That is, in the regime of applicability, the attenuation factor is constant for low values of magnetic field. In the next section, we emphasize the regimes where our results are valid and provide rough estimates for how the attenuation factor changes in other regimes. }

\section{Regimes of applicability}\label{ra}
There are several time and energy scales in this problem and we detail the regime of applicability of our results here. In this work, we have treated the sound wave as an externally applied field that modifies the band structure of the Weyl semimetals. The resulting electron-electron (e-e) relaxation has been analyzed through semi-classical  means using the Boltzmann equation. This method is only applicable for slow sound waves, that is, $\omega\tau^*\ll 1$. Further, we also expect that the wavelength of the sound wave is much larger than the mean free path of the electron, that is, $q\ell\ll 1$. If either or both of these conditions are not satisfied, then the sound wave can no longer be treated as a field \cite{AAA,Son}. Instead, its quantum mechanical nature becomes more relevant, and the system must be analyzed in terms of phonon absorption and scattering by electrons in the band. 

Further, we have evaluated the energy dissipation at low temperatures, which implies $\mu\gg k_BT$. We also treat the applied magnetic field to be weak enough that the Weyl cone does not undergo any significant Landau quantization. That is, $\mu\gg\hbar\omega_c$.  For larger magnetic fields, the system would be in the lowest Landau levels described by the chiral states in the Weyl semimetal, and there would be no Fermi surface geometry to speak of, rendering the geometric contribution to sound attenuation vanish. Instead, the only contribution to sound attenuation would be the $u_{zz}$ component that we have been ignoring so far \cite{Pikulin}. In a single cycle of the sound wave, let the chemical potential change from $\mu_1$ to $\mu_2$. The energy dissipated by this is given by
\begin{align}
    E=\int_{\mu_1}^{\mu_2}\epsilon D(\epsilon)d\epsilon \sim \frac{B}{\phi_0}\frac{(\Delta\mu)^2}{2} \,, 
\end{align}
where we have used $D(\epsilon)=\frac{B}{\phi_0}$ in the lowest Landau level. The number of effective cycles for a given relaxation time is $\omega\tau^*$, and the total energy dissipated can be estimated as $\omega\tau^*E$. The rate of change of energy dissipation is thus $\omega^2\tau^*E$. This approximate calculation yields 
\begin{align}
    \Gamma\sim\omega^2\tau^*\frac{B}{\phi_0}\frac{(\Delta\mu)^2}{2}/\frac{1}{2}\rho \omega^2u_0^2 \,. 
\end{align}
We can consider $\Delta\mu\sim|C|$, that is the magnitude of applied strain. The key takeaway here is that even in this case $\Gamma\sim\omega^2\tau^*$ like in the case of the geometric contribution. 
electron-electron (e-e)
We finally address the situation when the chemical potential approaches the Weyl nodes, that is, $\mu\rightarrow 0$. In this case the effective Fermi surface is a set of measure zero, and has no area whose geometry can be modified by strain.  Therefore, the geometric contribution obtained by e-e relaxation processes in this context would no longer contribute to sound attenuation. Interactions and collective phenomena like plasmons might become relevant in this regime. The lack of electronic states for relaxation in a non-interacting model can be rectified by allowing for donor impurities in the semimetal. The presence of donor impurity states creates a Fermi surface for the system to relax into. In 3D multi-Weyl semimetals, the donor impurity concentration $n_i=\int d\epsilon D(\epsilon) \sim \epsilon_F ^{1+2/n}$. In such a system, instead of the semi-classical approach, it is more appropriate to treat the relaxation time in terms of the Fermi golden rule\cite{Burkov, Park}. That is
\begin{align}
    \frac{1}{\tau_k}&=\int\frac{d^3k'}{(2\pi)^3}W_{kk'}(1-\cos\theta_{kk'})\\
    W_{kk'}&=\frac{2\pi}{\hbar}n_i|\langle V_{kk'}\rangle |^2\delta(\epsilon_k-\epsilon_{k'}) \,, 
\end{align}
where $\theta_{kk'}$ is the scattering angle from a state at $k$ to a state at $k'$. Here $V_{kk'}$ is the scattering potential due to the presence of impurities. This potential can be modeled in several ways, the simplest instances being those of short range scattering ($V_{kk'}=$constant) and Coulomb scattering ($V_{kk'}=\frac{4\pi e^2}{q^2+q_{TF}^2}$). Here $q=|k-k'|$ and the Thomas-Fermi screening length $q_{TF}^2\sim D(\epsilon)$. 

In the case of multi-Weyl semimetals, Fermi surface is not isotropic with respect to the axial plane ($k_x-k_y$) and the symmetry axis ($k_z$), which makes $q$ generally dependent on the specific $k$ and $k'$ as opposed to just $\theta_{kk'}$. A detailed analysis of the effects of this anisotropy is beyond the scope of the present work and has been studied in the existing literature in the context of conductance measurements. Instead, we make broader comments here as applicable to sound attenuation.

When the scattering potential is a constant, then it is straightforward to show that $\tau^{-1}_{k_i}\sim \epsilon^{2/n}$, that is, the scattering rate is proportional to the density of states. We can estimate the dependence of the rate of energy dissipated in such a system as
\begin{align}
    \langle \dot E\rangle \sim \omega^2\int_{\epsilon_F}^{\epsilon_F+\Delta\mu}   \epsilon\tau(\epsilon)  D(\epsilon) d\epsilon \sim \omega^2 \epsilon_{F} \Delta\mu \sim \omega^2 n_i^{\frac{1}{2}} \omega u_0/v_s \,. 
\end{align}
Therefore, we have $\Gamma\sim \omega n^{1/2}/\rho v_s u_0$. Similarly, when the system is in the strongly screened limit $q_{TF}\gg q$, so $V_{kk'}\equiv V(\epsilon)$, $\tau^{-1}_{k_i}\sim \epsilon^{2/n}/\epsilon_F^{4/n}$. Therefore, $\langle \dot E\rangle\sim \omega^2 \epsilon^3_{F} \Delta\mu \sim \omega^2 n_i^{\frac{3}{2}} \omega u_0/v_s $. This leads to  $\Gamma\sim \omega n^{3/2}/\rho v_s u_0$.

\section{Discussion}\label{disc}
In this work, we have shown that attenuation of ultrasonic sound in double-Weyl semimetals is fundamentally distinct from that of simple Weyl semimetals. This difference in relaxation mechanism in each case is a result of how strain couples to Weyl materials. The axial coupling of the strain to the two Weyl nodes, which is the dominant mode of relaxation due to sound in simple Weyl semimetals, is entirely absent in double Weyl semimetals. Instead, the change in the geometry of the Fermi surface due to the applied dynamic strain is the only contributing factor in double-Weyl semimetals, rendering it a measure of this geometry. 

\varsha{In higher multi-Weyl semimetals, the specific relaxation mechanism of relevance is determined by how strain couples to the Weyl nodes in question, though the Fermi surface would be deformed by strain for all winding numbers $n>1$. If the terms of the deformation potential $\lambda$ are purely even under inversion (for example, in materials with even $n$), then only geometric relaxation is possible. Whereas for materials where $\lambda$ is odd under inversion, then both forms of relaxation are possible. }

Additionally, we have specifically focused on planar sound waves in this study. This is because the deformation of the Fermi surface is induced only by the planar component of the strain. The only out-of-plane component of the sound wave that might couple to the Weyl material is the $u_{zz}$ term. This term acts like an axial vector, and its contributions to sound attenuation are much smaller in comparison to other possible mechanisms of relaxation \cite{Antebi}. Therefore, the geometric contribution would dominate sound attenuation even when non-planar sound is applied to the system. 

\varsha{Computational and experimental attempts to identify and characterize multi-Weyl semimetals have led to a few possible candidate materials, with HgCr$_2$Se$_4$ being a front-runner. While some synthetic materials have been proposed to realize the double-Weyl semimetal discussed in this work, the chief candidate material reported in the literature is HgCr$_2$Se$_4$ \cite{material1,material2,Singh2018,m4,m5,m3,material3}. We provide a rough estimate here of the sound attenuation factor $\Gamma$ in such materials. We assume a slow-moving sound wave with $\omega/2\pi\sim 10^2$ MHz and $v_s\sim 10^3$ m/s. We further assume $\tau^*\sim 10^{-11}$s and a doping such that $\mu\sim 2$meV. The density of the material is $6.350$ g/cm$^3$ and the unit cell volume $a^3=324.34 $ {\AA}$^3$ \cite{osti_1271332}. These parameters place the system in the right regime of validity of our results. Any true estimate will also depend crucially on the Gr\"{u}neisen parameter for which we use the typical value of $\gamma=19/6$ for the Lennard-Jones interatomic potential in 3D \cite{Krivtsov2011}. For an isotropic initial Weyl cone, it is also reasonable to state that the energy scales are such that $v_1\sim v_2$. Thus, we evaluate that magnitude of the attenuation factor in such a material to be of the order $\Gamma\sim 1.5$ kHz.

The key takeaway, however, is the non-trivial ways in which momentum space geometry affects physical properties of quantum materials.} The geometric tensor has garnered much interest in recent years as a potential tool in the identification and classification of band topology in correlated electronic systems. Its physical manifestation in experimentally accessible quantities continues to be explored in several systems. Here, we argue that since the geometric tensor acts as a metric over the Fermi surface, its effects are measurable in quantities that couple to deformations in the Fermi surface. Since dynamic strain (or in other words, sound waves) explicitly breaks the rotational symmetry of the Fermi surfaces of multi-Weyl semimetals, it offers a quantity that explicitly depends on this deformation, and thus the geometric tensor itself.

\section*{Acknowledgments}
The work was carried out under the auspices of the U.S. DOE NNSA under contract No. 89233218CNA000001 through the LDRD Program, and was supported by the Center for Nonlinear Studies at LANL (LYY), and was performed, in part, at the Center for Integrated Nanotechnologies, an Office of Science User Facility operated for the U.S. DOE Office of Science, under user proposals $\#2018BU0010$ and $\#2018BU0083$.

\bibliography{ref}
\clearpage
\onecolumngrid
\appendix
\section{Effect of strain on multi-Weyl semimetals}\label{A}

In this section, we provide a brief description of how strain couples to multi-Weyl semimetals for the sake of completeness. For further details, we direct the reader to our previous work in Ref. \cite{OldPaper}. The momentum space Hamiltonian for the double-Weyl semimetal is given by 
\begin{align}
\begin{split}
        H&=t_1(\cos{k_y}-\cos{k_x})\sigma_x+2t_1\sin{k_x}\sin{k_y}\sigma_y\\&+t_3\cos{k_z}\sigma_z+t_0(2-\cos{k_x}-\cos{k_y})\sigma_z \,.
\end{split}
\end{align}
The corresponding lattice Hamiltonian is
\begin{align}
\begin{split}
    H=&\sum_{i,j,k}\Big( -c_{i+1,j,k}^\dagger c_{i,j,k}(t_1\sigma_x+t_0\sigma_z) + t_3\sigma_z c_{i,j,k+1}^\dagger c_{i,j,k} \\&+ c_{i,j+1,k}^\dagger c_{i,j,k}(t_1\sigma_x-t_0\sigma_z)  + t_2\sigma_y c_{i+1,j+1,k}^\dagger c_{i,j,k}\\& - t_2\sigma_y c_{i+1,j-1,k}^\dagger c_{i,j,k} + 2t_0\sigma_z c_{i,j,k}^\dagger c_{i,j,k} + \textnormal{h.c.}\Big) \,.
\end{split}
\end{align}
Strain in the system can be incorporated into the Hamiltonian through the changes it induces in the hopping parameters. This idea has been explored in several studies before, and we employ it here in the following way:
\begin{align}
\begin{split}
        t_{\alpha\alpha}({\bf a_i}+\delta{\bf r_1})&=t_{\alpha\alpha}({\bf a_i})+\frac{{\bf a_i}\cdot\delta{\bf r_1}}{a_i}\frac{\partial{t_{\alpha\alpha}}}{\partial r}\Bigg|_{\bf a_i} \,, 
\end{split}\label{ts}
\end{align}
where $\alpha$ runs over the orbital degrees of freedom, $i$ runs over spatial directions and ${\bf a_i}$ are the unperturbed lattice vectors. On deforming the lattice thus, the modified low-energy Hamiltonian is now given by 
\begin{align}
\begin{split}
    H&=t_1\big(k_x^2-k_y^2-2\gamma(u_{xx}-u_{yy})\big)\sigma_x-2t_2(2k_xk_y-2\gamma_{xy})\sigma_y\\
    &+2t_3\big(k_z\pm \frac{\pi}{2}\pm\gamma u_{zz}\frac{\pi}{2}\big)\sigma_z+2\gamma t_0(u_{xx}+u_{yy})\sigma_z \,.
\end{split}
\end{align}
A slightly modified version of this Hamiltonian is used below and elsewhere in this work. As stressed in the main manuscript, it is clear that elements of the strain tensor do not couple minimally with the momentum terms in the Hamiltonian. That is, strain is neither a gauge field nor axial (since the coupling does not change sign between the nodes). A similar analysis can easily be performed for higher $n$ to show that strain is not a gauge field for any $n>1$.

It is also useful to see how strain modifies the overall energy band in the following way. We rewrite the low-energy Hamiltonian as 
\begin{align}
\begin{split}
        H&=v_1[(k_x^2-k_y^2)-C_x(x_i)]\sigma_x+v_1[2k_xk_y-C_y(x_i)]\sigma_y+v_2k_3\sigma_z\\&\equiv ({\bf Z}-{\bf C})\cdot{\bf \sigma}\label{Hs} \,. 
\end{split}
\end{align}
Thus, the energy bands are
\begin{align*}
   & E=\sqrt{[v_1(k_x^2-k_y^2)-C_x]^2+(2v_1k_xk_y-C_y)^2+v_2^2k_z^2}
   \\& \sim \sqrt{v_1^2(k_x^2+k_y^2)^2+v_2^2k_z^2-2C_xv_1(k_x^2-k_y^2)-2C_yv_1(2k_xk_y)}\\
    &\sim E_0 - {\bf C}\cdot\frac{{\bf Z}}{E_0}
    \equiv E_0+\delta E .
\end{align*}
One can cast the change in energy in terms of the quantum geometric tensor defined by 
\begin{equation}
g_{ab}(\mathbf{k}) = \Re\bigg({\langle \partial_{k_a} u_\mathbf{k}| \partial_{k_b} u_\mathbf{k}\rangle}-{\langle \partial_{k_a} u_\mathbf{k}|u_\mathbf{k}\rangle\langle u_\mathbf{k}| \partial_{k_b} u_\mathbf{k}\rangle}\bigg) \,,
\end{equation}
where $u_\mathbf{k}(\mathbf{r})$ is the periodic part of the Bloch function. For double-Weyl semimetals,
\begin{align}
\begin{split}
    g_{xx}&= 4v_1^2\frac{(k_x^2+k_y^2)}{E_0^4}[v_1^2(k_x^2+k_y^2)k_y^2+v_2^2k_z^2] \,,\\
    g_{yy}&= 4v_1^2\frac{(k_x^2+k_y^2)}{E_0^4}[v_1^2(k_x^2+k_y^2)k_x^2+v_2^2k_z^2] \,,\\
    g_{xy}&= -4v_1^4\frac{k_xk_y(k_x^2+k_y^2)^{2}}{E_0^4} \,.
\end{split}
\end{align}
Therefore, the change in energy is given by
\begin{align}
\begin{split}
    \delta &E=-v_1\bigg(\frac{C_x}{E_0}\frac{(g_{xx}-g_{yy})}{W_2^2}+\frac{C_y}{E_0}\frac{2g_{xy}}{W_2^2}\bigg)\\
    &=-v_1\bigg(\frac{(u_{xx}-u_{yy})}{E_0}\frac{(g_{xx}-g_{yy})}{W_2^2}+\frac{2u_{xy}}{E_0}\frac{2g_{xy}}{W_2^2}\bigg) \,,
    \end{split}
\end{align}
where $W_2=2v_1^2\frac{(k_x^2+k_y^2)}{E_0^2}.$ This expression easily generalizes to all multi-Weyl semimetals for winding $n>1$ in a straightforward manner. 

\section{Chiral kinetic theory in the presence of strain}\label{EOM}

In this section, we outline some of the steps in solving the Boltzmann equation in order to estimate various transport quantities associated with strained double-Weyl semimetals. 

The Boltzmann equation associated with the modified distribution function at one of the Weyl nodes is given by
\begin{align}
  \partial_t f_s+ \nabla_{\bf x}f_s\cdot{\bf \dot{x}} +\nabla_{\bf k}f_s\cdot{\bf \dot{k}} = C[f_s] \,, \label{BE}
\end{align}
where $C[f_s]$ is the collision integral associated with the node of winding $s=\pm$. To estimate the collision integral, we will use the method in Refs. \cite{CKT} and \cite{Dantas2018}, as described below. Let us first write down the equations of motion for the strained system.  The effective action obtained from diagonalizing the path integral of the strained double-Weyl semimetal is given by
\begin{align}
   S_{eff}= \int_{t_0}^{t_f}({\bf k}\cdot\dot{{\bf x}}+{\bf A}\cdot\dot{{\bf x}}-E \sigma_3-{\bf A^x}\cdot\dot{{\bf x}}-{\bf A^k}\cdot\dot{{\bf k}}) dt \,, 
\end{align}
where $E$ is the energy of the strained system described by Eq. (\ref{Hs}). Here, ${\bf A}$ is the vector potential due to electromagnetic fields, while ${\bf A^x}$ and ${\bf A^k}$ are emergent gauge fields in the system in the spatial and momentum sector, respectively. Thus, other than the electric and magnetic fields, we have three gauge-invariant quantities in this system:
\begin{align}
   {\bf \nabla}_x\times {\bf A^x}&={\bf b} \,,\label{bfield}\\
   {\bf \nabla}_k\times {\bf A^k}&={\bf \Omega} \,, \\
   S_{ij}&=\partial_{k_j}A^x_i-\partial_{x_i}A^k_j \,,
\end{align}
where $\Omega$ is the Berry curvature. The presence of a non-zero $S_{ij}$ implies an anisotropy in the system induced by strain. It is a cross-Berry curvature term consisting of mixed derivatives. Using these gauge invariant quantities, we can now write down the equations of motion from the effective action.
\begin{align}
    \dot{x}_i(\delta_{ij}+S^T_{ij})&=v_j+({\bf \dot{k}}\times {\bf\Omega})_j \,, \\
    \dot{k}_i(\delta_{ij}+S_{ij})&=E^t_j+({\bf \dot{x}}\times {\bf B^t})_j \,,
\end{align}
where ${\bf E^t}={\bf \tilde E}-{\bf \nabla_x}E+\partial_t{\bf A^x}$ and ${\bf B^t}={\bf B}-{\bf b}$ are the total modified fields. The modified phase space density $\Phi$ and Poisson brackets are given by
\begin{align}
    \Phi&=1+\Omega_iB^t_i+S_{ii} \,, \\
    \{x^i,k_j\}&=\frac{\delta^i_{j}+{B^t}i\Omega_j}{(1+\Omega\cdot B^t)}-\frac{S_{ij}^T}{(1+\Omega\cdot B^t)^2} \,.
\end{align}
The explicit expressions for these fields for convenient gauge choices are given below:
\begin{align*}
    {\bf A^x}&=\Bigg(-\frac{Z_x \partial_xC_y-Z_y \partial_xC_x}{Z(Z+Z_z)},-\frac{Z_x \partial_yC_y-Z_y \partial_yC_x}{Z(Z+Z_z)},0\Bigg),\\ {\bf A^k}&=2\Bigg(\frac{-Z_xk_y+Z_yk_x}{Z(Z+Z_z)},\frac{-Z_xk_x-Z_yk_y}{Z(Z+Z_z)},0\Bigg) \,,\\
    {\bf e}&=\partial_t{\bf A^x}-\nabla_{\bf x}E \,, \quad {\bf v}=-\partial_t{\bf A^k}+\nabla_{\bf k}E \,,\\ {\bf b}&=\nabla\times {\bf A^x}=\Bigg(0,0,\frac{Z_z}{Z}( \partial_xC_x \partial_yC_y- \partial_xC_y \partial_yC_x)\Bigg) \,,\\
    S_{ij}&=\partial_{k_j}A^x_i-\partial_{x_i}A^k_j \,,\quad Z=E=\sqrt{Z_x^2+Z_y^2+Z_z^2} \,,\\
    Z_x&=k_x^2-k_y^2- C_x,\quad Z_y=2k_xk_y- C_y,\quad Z_z=k_z \,.
\end{align*}

From the expressions of $A^k_i$ and $A^x_i$, we can now calculate $S_{ij}$. Since we have assumed no displacement in $z$ direction, and no $z$ dependence in elements of the strain tensor, $i$ and $j$ only go as $(x,y)$. Thus we have
\begin{align*}
     S_{xx}&=2\frac{k_y\partial_xC_x-k_x\partial_xC_y}{Z^2} \,, \\
     S_{xy}&=2\frac{k_x\partial_xC_x+k_y\partial_xC_y}{Z^2} \,, \\
     S_{yx}&=2\frac{k_y\partial_yC_x-k_x\partial_yC_y}{Z^2} \,, \\
     S_{yy}&=2\frac{k_x\partial_yC_x+k_y\partial_yC_y}{Z^2} \,. 
\end{align*}

We can solve the equations of motion to obtain a more explicit form (expanded up to two orders in strain)
\begin{align}
\begin{split}
    -(\delta_{ij}&+{\bf \Omega\cdot B^t}\delta_{ij}-{\bf \Omega\cdot B^t}S_{ij})\dot{k}_i=(\delta_{ij}-S_{ij})E_i^t-\epsilon_{jlk}v_lB^t_k+\epsilon_{jlk}S^T_{il}v_iB_k+S_{ij}\epsilon_{ilk}B_kv_l-S_{ij}\epsilon_{ilk}B_kS^T_{ml}v_m\\&-\Omega_jE_k^tB_k^t+S_{ij}\Omega_iB_kE_k^t+\Omega_jS_{ik}E_i^tB_k+\epsilon_{jlk}\epsilon_{imn}\Omega_nS_{il}^TB_k(\delta_{rm}-S_{rm})E_r^t\\&-S_{ij}\epsilon_{ilk}\epsilon_{qmn}S_{ql}^T\Omega_nB_k(\delta_{rm}-S_{rm})E_r^t \,, 
\end{split}\\
\begin{split}
    (\delta_{ij}&+{\bf \Omega\cdot B^t}\delta_{ij}-{\bf \Omega\cdot B^t}S^T_{ij})\dot{x}_i=(\delta_{ij}-S_{ij}^T)v_i+\epsilon_{jkl}E_l^t\Omega_k-\epsilon_{jlk}S_{il}E_i^t\Omega_k-S_{ij}^T\epsilon_{ilk}E_l^t\Omega_k+S_{ij}^T\epsilon_{ilk}S_{ql}\tilde E_q\Omega_k\\&+B_j^t\Omega_kv_k-B_jS^T_{mk}v_m\Omega_k-S_{ij}B_i\Omega_kv_k+S_{ij}B_i\Omega_kS_{lk}v_l-S_{il}\epsilon_{jlk}\epsilon_{imn}\Omega_kB_n(\delta_{qm}-S_{qm}^T)v_q\\&+S_{ij}^T\epsilon_{ilk}\epsilon_{qmn}S_{ql}\Omega_kB_n(\delta_{rm}-S_{rm}^T)v_r \,. 
\end{split}
\end{align}

\subsection{Outline for solving the Boltzmann equation}
To solve the Boltzmann equation, we first need to estimate the collision integral. Since the original Weyl node has split into two nodes of lower winding, let us label the associated electron distribution in each of the nodes $f_+$ and $f_-$. The bar on top indicates Fermi surface averages
\begin{align}
\begin{split}
    C[f_+]&=-\frac{f_+-\bar{f_+}}{\tau_{intra}}-\frac{f_+-\bar{f_-}}{\tau_{inter}}\\&=-(f_+-\bar{f_-})\bigg(\frac{1}{\tau_{inter}}+\frac{1}{\tau_{intra}}\bigg)-\frac{\bar{f_+}-\bar{f_-}}{\tau_{inter}}\\&=-\frac{f_+-\bar{f_-}}{\tau^*}+\Lambda \,. 
\end{split}
\end{align}
Formulating such a collision integral helps us estimate the modified distribution function as
\begin{align*}
&\bigg(\dot{E}+ {\bf \dot{x}}\cdot{\frac{\partial E}{\partial {\bf x}} }+{\bf \dot{k}}\cdot{\frac{\partial E}{\partial {\bf k}} }\bigg) \partial_E f_+=-\frac{f_+-\bar{f_+}}{\tau^*}+\Lambda \,, \\
    \implies f_+&=\bar{f_+}+\tau^*\Lambda - \tau^*\bigg(\dot{E}+ {\bf \dot{x}}\cdot{\frac{\partial E}{\partial {\bf x}} }+{\bf \dot{k}}\cdot{\frac{\partial E}{\partial {\bf k}} }\bigg) \partial_E f \,. 
\end{align*}

If $\Lambda(E)=-\frac{\bar{f_+}-\bar{f_-}}{\tau_{inter}}$, which is a constant over the Fermi surface, is non-zero, then there is an inter-node scattering process that can contribute to the relaxation processes. To estimate $\Lambda$, we can take a Fermi surface average over the entire Boltzmann equation in the following way: 
\begin{align}
\begin{split}
    &\overline{\partial_tf_++\nabla_{\bf x}f_+\cdot{\bf \dot{x}}+\nabla_{\bf k}f_+\cdot{\bf \dot{k}}} = -\overline{\frac{f_+-\bar{f_+}}{\tau^*}}+\Lambda(E)=\Lambda(E) \,. \label{l1}
\end{split}
\end{align}

Therefore, we have (up to two orders of strain) 
\begin{align}
    \begin{split}
       \Lambda&=\partial_Ef\overline{\dot{E}}+\partial_Ef\overline{(\delta_{ij}+{\bf \Omega\cdot B^t}\delta_{ij}-{\bf \Omega\cdot B^t}S_{ij})\dot{k}_i\partial_{k_j}E}+\partial_Ef\overline{(\delta_{ij}+{\bf \Omega\cdot B^t}\delta_{ij}-{\bf \Omega\cdot B^t}S^T_{ij})\dot{x}_i\partial_{x_j}E} \,. 
    \end{split}
\end{align}
After some tedious, but straightforward algebra, one obtains
\begin{align}
\begin{split}
\Lambda=&-s\frac{8\pi}{E^2}\big(\partial_xC_x\partial_x\dot{C_x}+\partial_yC_x\partial_y\dot{C_x}\big)\partial_E f + s\frac{2\pi}{3E^2}\big(\partial_x\dot{C_y}\partial_yC_x-\partial_x\dot{C_x}\partial_yC_y-\partial_y\dot{C_y}\partial_xC_x+\partial_y\dot{C_x}\partial_xC_y\big) \partial_Ef\\
&+s\frac{4\pi}{E}{\bf \tilde{E}\cdot B}\partial_Ef - s(\partial_xC_x\partial_yC_y-\partial_xC_y\partial_yC_x)\times \bigg(\frac{\pi^2}{E^4}(\tilde E_xB_x+\tilde E_yB_y)+\frac{3\pi^2}{E^4}\tilde E_zB_z\bigg)\partial_Ef \,. 
\end{split}\label{t1}
\end{align}

Thus, $\Lambda(E)$ gives us the amount of imbalance between the two nodes. Plugging this back into the Boltzmann equation gives us the full expression for $f_s$. Any response evaluated in this way will have to be summed over both nodes in the end.  

\end{document}